\documentstyle[12pt,epsf]{article}

\makeatletter

\def\section{\@startsection {section}{1}{\z@}{24pt plus 2pt minus 2pt}
{12pt plus 2pt minus 2pt}{\large\bf}}

\def\subsection{\@startsection {subsection}{2}{\z@}{12pt plus 2pt minus 2pt}
{12pt plus 2pt minus 2pt}{\normalsize\bf}}
\makeatother

\begin{document}

\date{}


\title{\Large\bf $N$-body Simulation of Galaxy Formation \\
on GRAPE-4 Special-Purpose Computer}

\author{Toshiyuki Fukushige and Junichiro Makino
\thanks{also, Department of Information Sciences and Graphics}\\
Department of General Systems Studies,\\ 
College of Arts and Sciences, University of Tokyo, \\ 
Tokyo 153, Japan \\ 
Email: fukushig@chianti.c.u-tokyo.ac.jp}

\maketitle

\subsection*{\centering Abstract} 

{\em We report on resent $N$-body simulations of galaxy formation
performed on the GRAPE-4 (\underbar{Gra}vity
\underbar{P}ip\underbar{e} 4) system, a special-purpose computer for
astrophysical $N$-body simulations. We review the astrophysical
motivation, the algorithm, the actual performance, and the price per
performance.  The performance obtained is 332 Gflops averaged over 185
hours for a simulation of a galaxy formation with 786,400 particles.
The price per performance obtained is 4,600 dollars per Gflops.  The
configuration used for the simulation consists of 1,269 pipeline
processors and has a peak speed of 663 Gflops.  }

\section{Introduction}

``How galaxies were formed ?" is one of the most important unsolved
problems in astrophysics. Various structures in the universe, such as
galaxies, are believed to be formed from small density fluctuation
through the gravitational instability.  The growth of instability in
linear region can be calculated analytically.  However, since the
present galaxy is in fully nonlinear region, most of the questions we
want to answer can be studied only by means of numerical simulations.
These questions include: ``when galaxies were formed, and which of
stars and galaxies were formed first?", ``what was building blocks of
galaxies?", ``what determined their morphology?, $i.e.,$ how
elliptical galaxies were formed ? and how spirals were formed ?", and
``how concentrated was the cores of galaxies at their birth and how
they evolved?"
 
The dominant force that drives the galaxy formation is the gravity,
although other effects such as energy dissipation through cooling of
gas and energy input from supernova explosion play important roles to
determine the details. 

The $N$-body simulation is the most straightforward technique to
follow the formation process.  Initial configuration of particles is
generated so that it expresses small density fluctuation in the early
universe. Then, we integrate the orbit of each particle in the
gravitational field made by the particles themselves.  We investigate
structural and dynamical properties of the simulated galaxies. 

The cosmological $N$-body simulation has been one of grand challenge
problems in computational physics. The 92' Gordon Bell prize is
awarded to $N$-body simulations by Warren and Salmon\cite{ws}. The
calculation cost of $N$-body simulation rapidly increases for large
$N$ because it is proportional to $N^2$. This is due to the fact that
the gravity is a long-range attractive force. In order to reduce
the calculation costs, some fast techniques, such as the particle-mesh (PM)
scheme, the particle-particle particle-mesh (P$^3$M) scheme, the
Barnes-Hut tree algorithm, have been used for this application. 

However, even the largest simulations, including the one by Warren and
Salmon, lack mass and spatial resolution to study properties of
individual galaxies.  These simulations are performed mainly to study
larger scale structures, such as galaxies clusters.  The simulation
volume is large (more than 10 mega persec scale).  Therefore, the
simulation volume contains many galaxies and the number of particles
available to express one galaxy is rather small (between a few hundred
and a few thousand).  Such low mass resolution makes it difficult to
study the structure and dynamics of individual galaxies.  Moreover,
the numerical error due to small $N$, such as two-body relaxation
effect, has considerable effects on the structure. 

In order to solve these difficulties, simulations that extract only a
small region around a peak of density fluctuation, $i.e.,$ seed of a
galaxy, have been performed.  For example, Dubinski and
Carlberg\cite{dubin} performed collapse simulations of density peak
using 32,000 particles in a 2 mega persec radius sphere. 

We report simulations of an isolated galaxy formation with much higher
mass and spatial resolutions. The number of particles is 786,400,
while the maximum number of particles in literature is 280,000 and
the typical number is 8,000-33,000. The spatial resolution is determined
by softening length of the gravitational force, which is introduced to
avoid numerical difficulties due to close encounters. The spatial
resolution of our simulation is 140 persec, which is ten times smaller
than the minimum value used in the previous simulations. 

The higher spatial and mass resolutions are necessary to investigate
finer structure of central part of galaxies.  Previous simulations
reported that the formed galaxies had no core and had $1/r$ density
cusp at a region just outside a limit of spatial resolution. Recently,
similar density cusp are observed in large elliptical galaxies by the
Hubble Space Telescope($e.g.$ \cite{lau}).  To determine whether the
simulation results is real or not, we have to perform simulations with
higher mass and spatial resolutions, $i.e.,$ with larger number of
particles and smaller softening length.  However, both cause a large
increase of the computing time. 

In order to perform simulations with much higher mass and spatial
resolutions, we used individual timestep algorithm\cite{aarseth} and a
special-purpose computer GRAPE-4\cite{SC94}\cite{SC95}.  The
individual timestep algorithm allows us to use smaller softening such
as 140 persec without large increase of the total number of timesteps,
and GRAPE-4 accelerates the individual timestep algorithm. 

In the simulation of a galaxy formation with higher spatial
resolution, the orbital timescale of particles ranges over many orders
of magnitudes. For example, the orbital timescale of stars in the core
of the galaxy is less than $10^6$ years, while that of stars in the
halo is about $10^8$ years. Thus, a small number of particles require
very short timesteps.  In the individual timestep scheme, each
particle has it own timestep $\Delta t_i$ and maintains its own time
$t_i$. To integrate the system, one first selects the particle for
which the next time ($t_i + \Delta t_i$) is the minimum. Then, one
predicts its position at this new time.  Positions of all other
particles at this time must be predicted also. Then the force on that
particle from other particles is calculated following Newton's law of
gravity. The position and velocity of the particle is then corrected. 

In order to accelerate astrophysical $N$-body simulation with
individual timestep algorithm, we developed a series of
special-purpose hardwares for the force calculation
\cite{Sugimoto1990}.  GRAPE has pipelines specialized for the
calculation of interactions of between particles, which is the most
expensive part of the $N$-body simulation.  Other calculations, such
as the time integration of orbits, are performed on the host computer
connected to GRAPE.  Figure 1 illustrates basic concept of the GRAPE
system.  In the simplest case, the host computer sends positions and
masses of all particles to GRAPE.  Then GRAPE calculates the forces
between particles, and sends them back to the host computer.  The
GRAPE system achieved high performance on the gravitational $N$-body
simulation through highly parallel, pipelined architecture specialized
for the force calculation. 

\begin{figure}
{\centering \leavevmode \epsfxsize=10cm \epsfbox{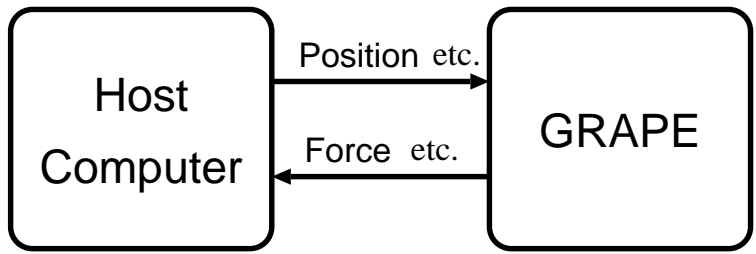}
\caption{Basic concept of GRAPE system}
}
\end{figure}

The rest of this paper is organized as follows.  In section 2, we
describe the GRAPE-4 system. In section 3 we present some calculation
results. In section 4 and 5, we report the performance and the price
per performance obtained on our simulation, respectively. In section
6, we briefly discuss on other calculations. 

\section{GRAPE-4 System}

We briefly describe architecture of the GRAPE-4 system.  GRAPE-4 is
designed to run the individual timestep algorithm with very high
speed.  More detailed descriptions of the GRAPE-4 system are in
\cite{SC94}.  We used 3 clusters configuration of the GRAPE-4 system,
which consists of a host computer and four clusters. One cluster has
one host-interface board (HIB), one control board (CB), and 9
processor boards (PB). HIB and CB handles the communication between
the processor boards and the host.  The processor boards perform the
force calculation.  Each processor board houses 48 HARP (Hermite
AccceleRator Pipeline) chips, which are custom LSI chips to calculate
the gravitational force and its first time derivative. 

A processor board consists of the particle data memory, one PROMETHEUS
LSI chip and 48 HARP LSI chips.  The particle data memory stores the
data of particles which exert the force. The PROMETHEUS LSI is used to
calculate the position (and velocity) of particles at a specified
time. This function is necessary to use the individual timestep.  The
HARP LSI chips calculate the gravitational accelerations and their
first time derivatives for particles.  Eight HARP chips are packaged
in one custom MCM package. One PB board houses 6 MCMs. We use 47 out
of 48 chips for actual calculation in order to utilize MCMs with one
defect chips. 

The theoretical peak speed is 663 Gflops.  Each clusters has 9
processor boards, and each processor board has 47 pipeline processors.
The total number of pipeline processors is 1269. Each processor
operates 49 floating operations in three clock cycles, and the clock
frequency is 32 MHz. 

\section{Calculation Result}

We present some results of our calculations. Here, we restrict it to
illustrate time evolution of particle distribution.  More
comprehensive analysis of the calculation result is presented elsewhere
\cite{fm96}.  We simulated formation of a galaxy in a 2 mega persec
radius sphere with 786,400 particles.  The individual particle each
represents about $4.0\times 10^6$ solar masses. The softening length
is 140 persec. 

We assigned the initial positions and velocities to particles in a
spherical region surrounding a density peak selected from a discrete
realization of density contrast field based on a standard Cold Dark
Matter scenario.  We performed numerical integrations of equation of
motion using the Aarseth scheme\cite{aarseth}.  The numerical
integration scheme includes the 4-th order order Hermite scheme
(\cite{ma92}) with the individual (hierarchical) timestep algorithm. 

Figures 2--5 show time evolution of a collapsing halo, and are
snapshots of particle distribution at redshift $z=8.7$, 5.1, 3.9, and
1.8, respectively.  The times corresponding to these redshift are
about 3.3\%, 6.6\%, 9.1\%, and 20\% of the age of the universe,
respectively. The length of the side of box is 240 kilo persec.
Figure 6 is as same as figure 5, but the length of the side of box is
24 kilo persec, which is 10 times smaller than that of figure 5. 

\section{Performance}

We report the performance statistics for one of recent simulations on
GRAPE-4. The performance numbers are based on the wall-clock time
obtained from UNIX system timer on the host computer (a DEC AXP 3900).
The total number of floating point operations is calculated as $49N
n$, where $N$ is the number of particles and $n$ is the number of
individual steps. Each pairwise interaction costs 49 flops, using the
Livermore Loops prescription of 1 square root = 1 division = 4 flops. 

We performed the simulation from $z=46$, where $z$ is redshift, to
$1.7$, for about $2.8\times 10^9$ years.  The total number of
individual steps was $5.735 \times 10^9$. The whole simulation took
$6.657\times 10^5$ seconds, resulting in the average computing speed
of 332 Gflops. 

\section{Price per Performance}

The total price of GRAPE-4 is 150 M JYE, of which 80 M JYE is spent
for the production of the hardware. Remaining 70 M JYE is spent for
the designing (including the cost of design software and workstations
to run the software).  Price per Gflops is 0.46 M JYE, which is, with
the present exchange rate of 1 dollar = 100 JYE, 4,600 dollars. 

\section{Comment on Other Scheme}

It is difficult to use fast algorithms, such as the tree algorithm and
P$^3$M scheme, for the simulation with higher resolution because of
the following two reasons. Firstly, it is difficult to implement such
fast force-calculation scheme with individual time algorithm.
McMillan and Aarseth \cite{McMillan1993} implemented high-accuracy
tree algorithm with individual timestep. For practical number of
particles, however, this code turned out to be slower than the direct
method on 1-processor Cray YMP.  Secondly, in such fast scheme the
force is calculated with rather low accuracy. It is not clear whether
we can follow the evolution of a galaxy with highly concentrated core
using the calculated force with such low accuracy. 

The implementation of the individual timestep algorithm on massively
parallel processors is nontrivial, since the communication between
nodes must be extremely fast in order to use large number of
processors efficiently. The effort to implement the individual
timestep algorithm on a Cray T3D has not been very successful (Rainer
Spurzem and Douglas Heggie, private communication).  An implementation
which achieved a reasonable performance (several hundred Mflops) on
CM-2 with 256 FPUs was reported recently.  However, this algorithm is
not scalable to larger number of processors. In addition, it requires
an extremely fast broadcast mechanism. 

\section*{Acknowledgment}

We are grateful to Makoto Taiji for developing the GRAPE-4 system.  We
also thank Daiichiro Sugimoto and Toshikazu Ebisuzaki who have been
leading the GRAPE project from the start.  To generate initial
condition, we use the COSMIC package\cite{bert} developed by Edmund
Bertschinger, to whom we express our thanks.  This work was supported
by the Grant-in-aid for Specially Promoted Research (04102002) of the
Ministry of Education, Science, and Culture.

\section*{Biographical Sketch}

{\bf Toshiyuki Fukushige} received the B.S., M.S., and Ph. D. degrees
in systems science from University of Tokyo in 1991, 1993, and 1996,
respectively. In 1996 he was a postdoctoral fellow of Japan Society for
the Promotion of Society at University of Tokyo.  Since 1996 he has
been an research associate in the Department of General Systems Studies
of College of Arts and Science at University of Tokyo.  His Email
address is fukushig@chianti.c.u-tokyo.ac.jp and his home page is
http://grape.c.u-tokyo.ac.jp/pub/people/fukushig . 

\bigskip

{\bf Junichiro Makino} received the B.S., M.S., and Ph. D. degrees in
systems science from University of Tokyo in 1985, 1987, and 1990,
respectively. From 1990 to 1994 he was a research associate at
University of Tokyo. Since 1994 he has been an associate professor in
the Department of Information Science and Graphics of College of Arts
and Science at University of Tokyo.  He is one of the winners of 1995
Gordon Bell Prize for Special-Purpose computer.  His Email address is
makino@chianti.c.u-tokyo.ac.jp and his home page is
http://grape.c.u-tokyo.ac.jp/pub/people/makino/index-e.html .

\begin{figure}
{\centering \leavevmode \epsfxsize=4in \epsfbox{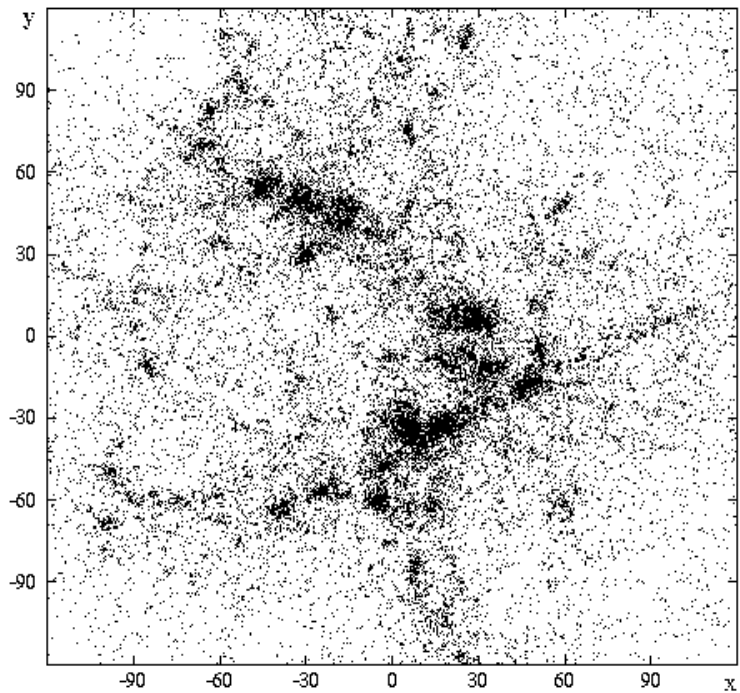}
\caption{Snapshot of particle distribution at redshift
$z=8.7$ (3.3\% of the age of the universe). The box
is 240 kilo persec wide. The number of particles is 786,400. The mass and
spatial resolutions are about $4.0\times 10^6$ solar masses and 140
persec, respectively.}
}
\end{figure}

\begin{figure}
{\centering \leavevmode \epsfxsize=4in \epsfbox{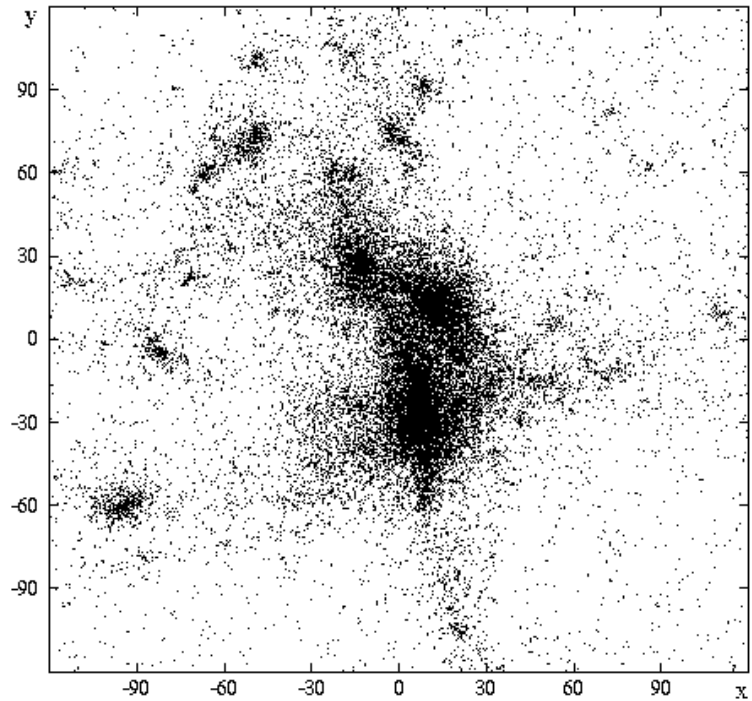}
\caption{Same as figure 2, but at $z=5.1$ (6.6\% of the age of the universe).}
}
\end{figure}

\begin{figure}
{\centering \leavevmode \epsfxsize=4in \epsfbox{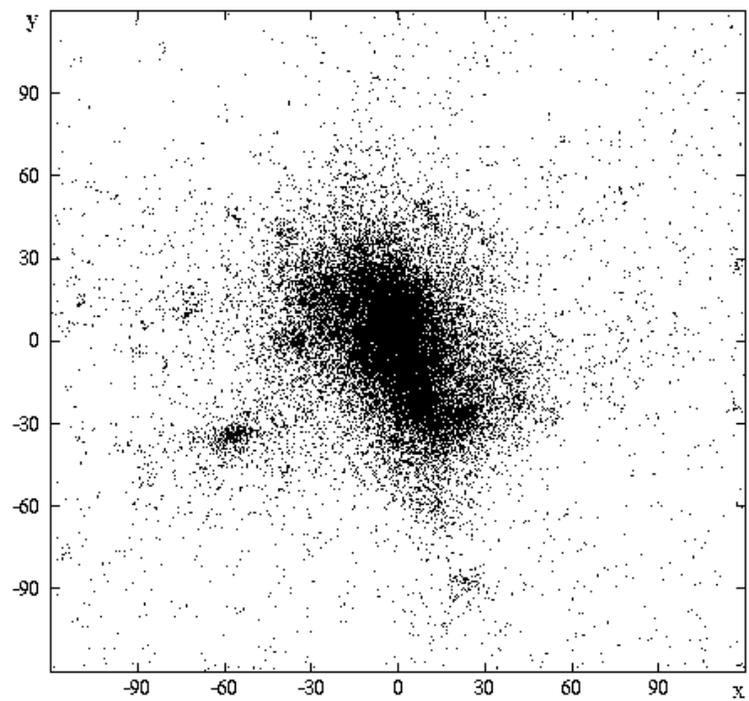}
\caption{Same as figure 2, but at $z=3.9$ (9.1\% of the age of the universe).}
}
\end{figure}

\begin{figure}
{\centering \leavevmode \epsfxsize=4in \epsfbox{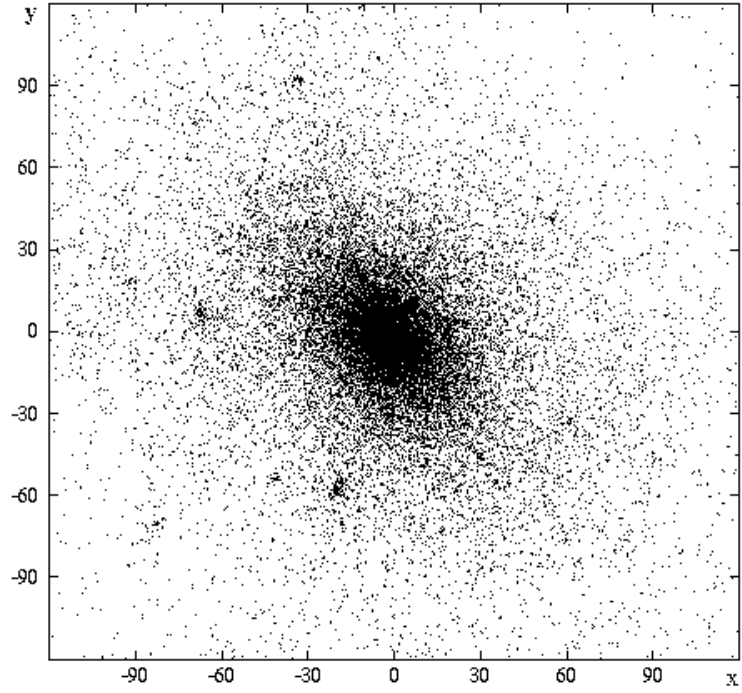}
\caption{Same as figure 2, but at $z=1.8$ (20\% of the age of the universe).}
}
\end{figure}

\begin{figure}
{\centering \leavevmode \epsfxsize=4in \epsfbox{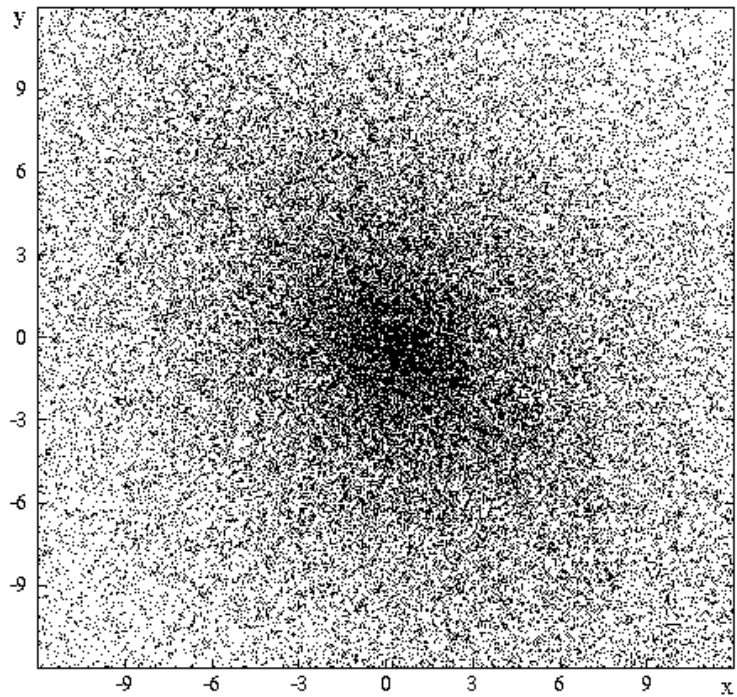}
\caption{Same as figure 5, but the box is 24 kilo persec wide.}
}
\end{figure}

\end{document}